\documentstyle[fleqn,twoside,gc]{article}
\def\ii{\'{\i}}
\def\bi{\bigskip}
\def\noi{\noindent}
\def\be{\begin{equation}}
\def\en{\end{equation}}
\def\bq{\begin{eqnarray}}
\def\eq{\end{eqnarray}}
\begin{document}

\twocolumn[
\Title{{\bf SUSY Cosmological Models}}

\bi

\Authors{F. Aceves de la Cruz\foom 1,J.J. Rosales\foom 2}{V.I. Tkach\foom3,J.
 Torres A.\foom4}{Instituto de F\ii sica, Universidad de Guanajuato\\ 
Lomas del Bosque 103, Lomas del Campestre\\ 37150 Le\'on, Guanajuato, M\'exico}
{Instituto de F\ii sica, Universidad de Guanajuato\\ 
Lomas del Bosque 103, Lomas del Campestre\\ 37150 Le\'on, Guanajuato M\'exico}
\Abstract{In this work we consider the action for a set of complex scalar 
supermultiplets interacting with the scale factor in the supersymmetric 
cosmological models. We show that the local conformal supersymmetry 
leads to a scalar field potential defined in terms of the K\"ahler potential 
and superpotential. Using supersymmetry breaking, we are able to obtain a     
normalizable wavefunction for the FRW cosmological model.}
]

\email 1 {fermin@ifug2.ugto.mx}
\email 2 {rosales@ifug3.ugto.mx}
\email 3 {vladimir@ifug3.ugto.mx}
\email 4 {jtorres@ifug1.ugto.mx}

\section{Introduction}
The study of supersymmetric minisuperspace models has led to important    
and interesting results. To find the physical states, it is sufficient   
to solve the Lorentz and supersymmetric constraints \cite{1,2,3}. 
Some of these results have already been presented in two comprehensive 
and organized works: a book \cite{4} and an extended review \cite{5}. 
In previous works \cite{6,7} we have proposed a new approach to the  
study of supersymmetric quantum cosmology. The main idea is to extend the
group of local time reparametrization of the cosmological mo-dels to the
$n=2$ local conformal time supersymmetry. For this purpose the odd ``time"
parameters $\eta , \bar\eta$ were introduced (where $\bar\eta$ is the complex
conjugate to $\eta$), which are the superpartners of the usual time 
para-meters. The new functions, which previously were functions of time $t$ 
become now superfunctions depending on $(t, \eta ,\bar\eta )$, which are 
called superfields. Following the superfield procedure we have constructed the 
superfield action for the cosmological models, which is invariant under 
$n=2$ local conformal time supersymmetry.\\ 
The fermionic superpartners of the scale factor and the homogeneous scalar
fields at the quantum level are elements of the Clifford algebra.\\
We will consider the supersymmetric FRW model interacting with a set of $n$
complex homogeneous scalar supermatter fields. We show that in this
case, the potential of scalar matter fields is a function of the K\"ahler
function and an arbitrary parameter $\alpha$. The local conformal
supersymmetry cannot fix the value of the parameter $\alpha$, the
space-time supersymmetry does. Furthermore, when $\alpha=1$, the scalar
field potential becomes the vacuum energy of the scalar fields interac-ting
with the chiral matter multiplets as in the case of $N=1$ supergravity
theory, \cite{8}. Using supersymmetry we are able to obtain a wavefunction 
which depende of the K\"ahler function.

\bi

\section{Supersymmetric FRW model with matter fields}
Let us begin by considering the FRW action

\[
S_{grav} =\frac{6}{8\pi G_N} \int \left(-\frac{R\dot R^2}{2N} +\frac{1}{2}
k NR\right.
\]
\be
\left. + \frac{d}{dt} \left(\frac{R^2\dot R}{2N}\right)\right) dt ,
\en
 \noi where $k = 1,0,-1$ stands for a spherical, plane and hyperspherical
three-space, respectively, $\dot R =\frac{dR}{dt}, G_N$ is the Newtonian
gravitational constant, $N(t)$ is the lapse function and $R(t)$ is the 
scale factor depending only on $t$. In this work we shall set $c=\hbar =1$.

\bi

\noi It is well known that the action (1) preserves the invariance under
the time reparametrization.

\bi

\be
t^\prime \to t + a (t)
\en

\noi if $R(t)$ and $N(t)$ are transformed as
\be
\delta R = a \dot R, \qquad \delta N = (aN)^{\bf .} .
\en
In order to obtain the superfield formulation of the action (1), the 
transformation of the time reparametrization (2) were extended to the $n=2$ 
local conformal time supersymmetry $(t, \eta ,\bar\eta )$ \cite{6}. 
These transformations can be written as 
\[
\delta t = {I\!\! L} (t,\eta , \bar\eta )+
\frac{1}{2} \bar\eta D_{\bar\eta}
{I\!\!L} (t,\eta , \bar \eta )
\]
\[
- \frac{1}{2} \eta D_\eta {I\!\!L}
(t,\eta ,\bar\eta ).
\]
\be
\delta \eta = \frac{i}{2} D_{\bar\eta} {I\!\!L} (t,\eta , \bar\eta ), \\
\delta\bar\eta = -\frac{i}{2}D_\eta {I\!\!L}(t,\eta ,\bar\eta ),\nonumber
\en 

\noi with the superfunction ${I\!\!L} (t,\eta ,\bar\eta )$, defined by
\be
{I\!\!L} (t, \eta ,\bar\eta ) =a (t)+i \eta \bar\beta^\prime (t) + i\bar\eta
\beta^\prime (t) + b(t) \eta\bar\eta ,
\en

\noi where $D_\eta =\frac{\partial}{\partial\eta} +i\bar\eta
\frac{\partial}{\partial t}$ and $D_{\bar\eta} =-
\frac{\partial}{\partial \bar\eta} -i\eta \frac{\partial}{\partial t}$ are 
the supercovariant derivatives of the global conformal supersymmetry with 
dimension $ [D_\eta] = l^{-1/2}$, $a(t)$ is a local time reparametrization 
parameter, $\beta^\prime (t) =N^{-1/2} \beta (t)$ is the Grassmann complex 
parameter of the local conformal SUSY transformations (4) and $b(t)$ is the 
parameter of local $U(1)$ rotations on the complex coordinate $\eta$.

\bi

The superfield generalization of the action (1), which is invariant under
the transformations (4), was found in our previous work \cite{6} and
it has the form

 \[
S_{grav} = \frac{6}{\kappa^2} \int \left\{ -\frac{{I\!\!N}^{-1}}{2} {I\!\!R}
D_{\bar\eta}{I\!\!R} D_\eta {I\!\!R}+\frac{\sqrt{k}}{2}{I\!\!R}^2\right.
\]
\[
 +\frac{1}{4}
D_{\bar\eta}({I\!\!N}^{-1}{I\!\!R}^2D_\eta {I\!\!R}) 
\]
\be
\left. - \frac{1}{4} D_\eta ({I\!\!N}^{-1} {I\!\!R}^2 D_{\bar\eta} 
{I\!\!R}) \right\} d\eta d\bar\eta dt , 
\en

\noi where we introduce the parameter $\kappa^2 = 8\pi G_N$. We can also see
that this action is hermitian for $k=0,1$. The last two terms in (6) form a 
total derivative which are necessary when we consider interaction.
${I\!\!N} (t, \eta ,\bar\eta )$ is a real one-dimensional gravity superfield
which has the form

\be
{I\!\!N} (t,\eta ,\bar\eta )= N(t)+i\eta\bar\psi^\prime (t) +i\bar\eta
\psi^\prime (t) + \eta\bar\eta V^\prime (t) ,
\en 

\noi where $\psi^\prime (t) =N^{1/2} \psi (t), \bar\psi^\prime (t) =N^{1/2}
\bar\psi (t)$ and $V^\prime (t)=NV +\bar\psi\psi$. This superfield transforms
as

\be
\delta {I\!\!N} = ({I\!\!L}{I\!\!N})+\frac{i}{2} D_\eta {I\!\!L} D_\eta
{I\!\!N} + \frac{i}{2} D_\eta {I\!\!L} D_\eta {I\!\!N} .
\en

The components of the superfield ${I\!\!N} (t,\eta ,\bar\eta )$ in (7) are 
the gauge field of the one-dimensional n=2 extended supergravity.

The superfield ${I\!\!R} (t,\eta ,\bar\eta )$ may be written as

\be
{I\!\!R} (t,\eta ,\bar\eta ) =R(t)+i\eta \bar\lambda^\prime (t) + i\bar\eta
\lambda^\prime (t) + \eta\bar\eta B^\prime (t) ,
\en

\noi where $\lambda^\prime (t)=\frac{\kappa N^{1/2}}{\sqrt{R}} \lambda (t),
\bar\lambda^\prime (t) =\frac{\kappa N^{1/2}}{\sqrt{R}}\bar\lambda (t)$
and $B^\prime (t) =\kappa NB -\frac{\kappa}{6\sqrt{R}} (\bar\psi\lambda -
\psi\bar\lambda )$. The transformation rule for the real scalar superfield 
${I\!\!R} (t,\eta ,\bar\eta )$ is

\be
\delta {I\!\!R} = {I\!\!L}\dot {I\!\!R}+\frac{i}{2} D_{\bar\eta} {I\!\!L}
D_\eta {I\!\!R} + \frac{i}{2} D_\eta {I\!\!L} D_{\bar\eta} {I\!\!R} .
\en

The component $B(t)$ in (9) is an auxiliary degree of freedom; $\lambda (t)$
and $\bar\lambda (t)$ are the fermionic superpartners of the scale factor
$R(t)$. The superfield transformations (8), (10) are the generalization of 
the transformations for $N(t)$ and $R(t)$ in (3).

\bi

The complex matter supermultiplets $Z^A (t,\eta ,\bar\eta )$ and
$\bar Z^{\bar A} (t,\eta , \bar\eta )= (Z^A)^{\dagger}$ consist of a set of 
spatially homogeneous matter fields $z^A (t)$ and $\bar z^{\bar A}
(t) (A=1,2,\ldots , n)$, four fermionic degrees of freedom $\chi^A (t),
\bar\chi^{\bar A} (t), \phi^A (t)$ and $\bar\phi^{\bar A} (t)$, as well as
the bosonic auxiliary fields $F^A (t)$ and $\bar F^{\bar A} (t)$.

\bi

The components of the matter superfields\\ 
$Z^A (t,\eta ,\bar\eta )$ and
$\bar Z^{\bar A} (t,\eta ,\bar \eta )$ may be written as 

\be
Z^A=z^A (t) +i\eta \chi^{\prime A} (t) +i\bar\eta\phi^{\prime A} (t) +
F^{\prime A} (t) \eta\bar\eta ,
\en

\be
\bar Z^{\bar A}=\bar z^A (t) +i\eta \bar\phi^{\prime \bar A} (t) +i\bar\eta
\bar\chi^{\prime \bar A} (t) + \bar F^{\prime \bar A} (t) \eta\bar\eta ,
\en

\noi where

\[ \chi^{\prime A} (t) = N^{1/2} R^{-3/2} \chi^A (t), \]
\[
\phi^{\prime A} (t) = N^{1/2} R^{-3/2} \phi^A (t), \]

\[ F^{\prime A}(t)=NF^A-\frac{1}{2}R^{-3/2}(\psi\chi^A -\bar\psi\phi^A). \]
 
\bi

The transformation rule for the superfields $Z^A (t, \eta ,\bar\eta)$ and
$\bar Z^{\bar A} (t,\eta ,\bar\eta )$ may be written as
\be
\delta Z^A= {I\!\!L}\dot Z^A +\frac{i}{2} D_\eta {I\!\!L} D_\eta Z^A +
\frac{i}{2} D_\eta {I\!\!L} D_\eta Z^A ,
\en

\be
\delta \bar Z^{\bar A}= {I\!\!L}{\dot{\bar{Z}}}^{\bar A} +\frac{i}{2} D_\eta
{I\!\!L} D_\eta \bar Z^{\bar A}+\frac{i}{2} D_{\bar \eta} {I\!\!L} D_\eta 
\bar Z^{\bar A}.
\en

So, the superfield action takes the form

\[
S =\int \left\{ - \frac{3}{\kappa^2} {I\!\!N}^{-1} {I\!\!R} D_{\bar{\eta}}
{I\!\!R} D_\eta {I\!\!R} + \frac{3}{\kappa^2} \sqrt{k} {I\!\!R}^2\right.
\]
\[
-\frac{2}{\kappa^3} {I\!\!R}^3 e^{\frac{ G}{2}} +
 \frac{1}{2\kappa^2}N^{-1}{I\!\!R}^3 G_{\bar AB}
\]
\be
\left.\left[ D_{\bar\eta}
\bar Z^{\bar A} D_\eta Z^B + D_{\bar\eta} Z^B D_\eta \bar Z^{\bar A} \right] 
\right\} d\eta d\bar\eta dt ,
\en

where $\kappa^2 = 8\pi{G_N}$. The action (15) is defined in terms of one 
arbitrary K\"ahler superfunction $G(Z^A, \bar Z^{\bar A})$ which is a
special combination of ${I\!\!K} (Z^A, \bar Z^{\bar A})$ and
$g(Z^A)$, i.e.

\be
G (Z,\bar Z) = {I\!\!K} (Z,\bar Z)+ \log |g (Z)|^2.
\en

\noi and is invariant under the transformations

\bq
g(Z) &\to& g (Z) \exp f (Z), \nonumber \\
{I\!\!K} (Z,\bar Z) &\to& {I\!\!K} (Z,\bar Z)-f (Z) -\bar f (\bar Z) ,
\eq

\noi with the K\"ahler potential ${I\!\!K} (Z,\bar Z)$ defined by the complex
superfield $Z^A$ related to the $G(Z,\bar Z)$ from (16). The superfunction
$G(Z,\bar Z)$ and their transformations are the generalizations of the 
K\"ahler function $G(z,\bar z)={I\!\!K} (z,\bar z)+ \log |g(z)|^2$ defined 
on the complex manifold. 

~Derivatives ~of ~K\"ahler ~function ~are ~denoted ~by ~$\frac{\partial
G}{\partial_{z^A}}=G_{,A}\equiv G_A, ~\frac{\partial G}{\partial_{{\bar
z}^{\bar A}}} = G_{,\bar A} \equiv G_{\bar A}, \frac{\partial^n
G}{\partial z^A \partial z^B \partial \bar z^{\bar C} \ldots \partial \bar
z^{\bar D}} =G_{,AB\bar C \ldots \bar D} \equiv G_{AB\bar C\ldots \bar D}$
and the K\"ahler metric is $G_{A\bar B}= G_{\bar BA}= K_{A\bar B}$, the
inverse K\"ahler metric $G^{A\bar B}$, such as $G^{A\bar B} G_{\bar BD}
=\delta^A_D$ can be used to define $G^A \equiv G^{A\bar B} G_{\bar B}$ and
$G^{\bar B} \equiv G_A G^{A\bar B}$. The action (15) is invariant under
the local $n = 2$ conformal supersymmetry transformations (4) if the
superfields are transformed as (8), (10), (13) and (14). The action (15)
corresponds to FRW in the minisuperspace sector of supergravity coupled to
complex scalar fields \cite{8}. After the integration over the
Grassmann variables $\eta ,\bar \eta$ the action (15) becomes a component
action with the auxiliary fields $B(t) , F^A (t)$ and $\bar F^{\bar A}
(t)$. These fields may be determined from the component action by taking
the variation with respect to them. The equations for these fields are
algebraical and their solutions are

\[
B=-\frac{\kappa}{18R^2} \bar\lambda\lambda +\frac{\sqrt{k}}{\kappa}
+\frac{1}{4\kappa R^2} G_{\bar AB} (\bar \chi^{\bar A} \chi^B 
+ \phi^B \bar\phi^{\bar A}) 
\]
\[
- \frac{R}{\kappa^2} e^{G/2} ,
\]

\[
F^D =-\frac{\kappa}{2R^3} (\bar\lambda \phi^D -\lambda\chi^D )-\frac{1}{R^3}
G^{D\bar A} G_{\bar ABC} \chi^C \phi^B 
\]
\[
+\frac{2}{\kappa} G^{D\bar A} 
(e^{G/2})_{,\bar A} .
\]    

After substituting them again into the component action we get the 
fo\-llo\-wing action:

\[
S =\int \left\{ -\frac{3}{\kappa^2} \frac{R(DR)^2}{N}-NR^3U (R,z,\bar z)
+\frac{2i}{3} \bar\lambda D\lambda\right.
\] 
\[
+ \frac{N\sqrt{k}}{3R} \bar\lambda\lambda
-\frac{N}{\kappa} e^{G/2} \bar\lambda\lambda 
 + \frac{\sqrt{k}}{\kappa}\sqrt{R}(\bar\psi\lambda -\psi\bar\lambda )
\]
\[
+\frac{R^3}{N\kappa^2} G_{\bar AB} D\bar z^{\bar A} D z^B +\frac{i}{2\kappa}
D z^B (\bar\lambda G_{\bar AB}\bar\chi^{\bar A}+\lambda G_{\bar AB}
\bar\phi^{\bar A} ) 
\]
\[
+ \frac{i}{2\kappa} D \bar z^{\bar A} (\bar\lambda G_{\bar AB} 
\phi^B +\lambda G_{\bar AB} \chi^B)-\frac{i}{\kappa^2} 
\]
\[
G_{\bar AB}
(\bar \chi^{\bar A} \tilde D \chi^B
 + \bar\phi^{\bar A} \tilde D \phi^B ) -
 \frac{N}{\kappa^2 R^3} R_{\bar AB\bar CD}\bar\chi^{\bar A}\chi^B
\bar\phi^{\bar C} \phi^D 
\]
\[
-\frac{i}{4\kappa\sqrt{R^3}} (\psi\bar\lambda -
\bar\psi\lambda ) G_{\bar AB}
 (\bar\chi^{\bar A} \chi^B + \phi^B
\bar\phi^{\bar A}) 
+ \frac{3N}{16\kappa^2 R^3}
\]
\[
 \left[ G_{\bar AB} (\bar\chi^{\bar A}\chi^B +\phi^B \bar\phi^{\bar A})
\right]^2 
+\frac{3\sqrt{k}}{2\kappa^2 R}
\]
\[
G_{\bar AB} 
(\bar\chi^{\bar A} \chi^B 
+ \phi^B \bar\phi^{\bar A})
- \frac{3N}{2\kappa^3} e^{G/2} G_{\bar AB} (\bar\chi^{\bar A}\chi^B
+ \phi^B \bar\phi^{\bar A})
\]
\[
- \frac{2N}{\kappa^3} (e^{G/2})_{,AB} \chi^A
\phi^B 
- \frac{2}{\kappa^3} N (e^{G/2})_{,\bar A \bar B} \bar\phi^{\bar A}
\bar\chi^{\bar B} 
\]
\[
-\frac{2}{\kappa^3} N(e^{G/2})_{,\bar AB}
 (\bar\chi^{\bar A}
\chi^B + \phi^B \bar\phi^{\bar A}) -
 \frac{N}{\kappa^2} \bar\lambda \left[ (e^{G/2})_{,A} \phi^A\right. 
\]
\[
\left.+(e^{G/2})_{,\bar A} \bar\chi^{\bar A}\right] +
 \frac{N}{\kappa^2} \lambda \left[ (e^{G/2})_{,A}\chi^A +
(e^{G/2})_{,\bar A} \bar\phi^{\bar A} \right] 
\]
\[
-\frac{\sqrt{R^3}}{\kappa^2}
(\bar\psi\lambda -\psi\bar\lambda ) e^{G/2} +
 \frac{\sqrt{R^3}}{\kappa^3} (e^{G/2})_{,A} (\psi\chi^A -\bar\psi
\phi^A)
\]
\be
\left.+ \frac{\sqrt{R^3}}{\kappa^3} (e^{G/2})_{,\bar A}
 (\psi\bar\phi^{\bar A} -
\bar\psi\bar\chi^{\bar A}) \right\} dt ,
\en  

\noi where $DR=\dot R-\frac{\kappa}{6\sqrt{R}}(\bar\psi\lambda
+\psi\bar\lambda ), D_{z^A} =\dot z^A-\frac{i}{2\sqrt{R^3}} (\bar\psi\phi^A
+\psi\chi^A), D \chi^B =\dot\chi^B -\frac{i}{2} V\chi^B$,
$D\phi^B=\dot\phi^B+\frac{i}{2} V\phi^B$, $D\lambda=\dot\lambda
+\frac{i}{2}V\lambda$, $\tilde D\chi^B = D\chi^B +\Gamma^B_{CD} \dot z^C
\chi^D$, $\tilde D \phi^B=D\phi^B + \Gamma^B_{CD} \dot z^C \phi^D$,
$R_{\bar AB\bar CD}$ is the curvature tensor of the K\"ahler manifold
defined by the coordinates $z^A$, $\bar z^{\bar B}$ with the metric
$G_{A\bar B}$, and $\Gamma^B_{CD}=G^{B\bar A} G_{\bar ACD}$ are the
Christoffel symbols in the definition of the covariant derivatives and
their complex conjugate. The kinetic energy term of the scalar factor
$R(t)$ is not positive in the action, (1), (6), (15) and (18), as is
usually the case, but negative. This is due to the fact that the
particle-like fluctuations do not correspond to the scalar factor $R(t)$
\cite{9}. Besides, the potential term $U(R,z,\bar z)$ reads

\be
U(R,z,\bar z) =-\frac{3k}{\kappa^2 R^2} +\frac{6\sqrt{k}}{\kappa^3 R}
e^{G/2} +V_{eff}(z,\bar z) ,
\en
\noi where the effective potential of the scalar matter fields is

\[
V_{eff}=\frac{4}{\kappa^4} \left[ (e^{G/2})_{,\bar A} G^{\bar AD}
(e^{G/2})_{,D} -\frac{3}{4} e^G \right] 
\]
\be
=\frac{e^G}{\kappa^4} [G^A
G_A -3].
\en

In the action (18), as in the effective potential, the K\"ahler function
is a function of scalar fields $G(z,\bar z)$. From (19) we can see that 
when $k=0$, $U(R,z,\bar z) = V_{eff} (z,\bar z)$.

\bi

In order to discuss the implications of spontaneous supersymmetry 
breaking we need to display the potential (20) in terms of the auxiliary 
fields

\be
V_{eff} (z,\bar z) =\frac{\bar F^{\bar A} G_{\bar AB} F^B}{\kappa^2} -
\frac{3B^2}{R^2} ,
\en

\noi where the auxuliary fields $B$ and $F^A$ now read

\be
B=\frac{R}{\kappa^2} e^{G/2} ,
\en

\be
F^A = \frac{1}{\kappa}e^{G/2} G^A .
\en

\bi

The supersymmetry is spontaneously broken, if the auxiliary fields (23) 
of the
matter supermultiplets get nonvanishing vacuum expectation values. The 
potential (20,21) consists of two terms; the first of them is the potential 
for the scalar fields in the case of global supersymmetry. Indeed this 
superpotential is not positive semi-definite in contrast with the standard 
supersymmetric quantum mechanics case. The global supersymmetry \cite{10} 
is unbroken when the energy is zero due to $F^A =0$. Besides,
the energy plays the role of the order parameter in this case. For the local 
symmetry, the energy ceases to play the role of the order parameter when
gravity is taken into account \cite{8} in other words, the spontaneous
breaking of supersymmetry in our model, allows us to describe the general
physical situation for different energies, including the case when the energy
is zero.

\bi

Now we can see that at the minimum in (21) $V_{eff} (z^A_0,$
$ \bar z^{\bar A}_0)=0$, but $F^A\not = 0$, then the supersymmetry is broken
 when the vacuum energy is zero. The measure of this breakdown is the term
$(-\frac{1}{\kappa} e^{2G (z^A, \bar z^{\bar A})})\bar\lambda\lambda$ in the
action (18). Besides, we can identify

\be
m_{3/2}=\frac{1}{\kappa} e^{\frac{G}{2} (z^A_0, \bar z^{\bar A}_0)},
\en

\noi as the gravitino mass in the effective supergravity theory \cite{8}.
Hence, we can see that in our model the conformal time supersymmetry (4), 
being a subgroup of the space-time SUSY, gives us a mechanism of spontaneous
breaking of this SUSY \cite{8}.

\bi

\section{Wave function of the Universe}
The Grassmann components of the vacuum configuration with the FRW metric
may be obtained by decomposition of the Rarita-Schwinger field and of the
spinor field in the following way \cite{11} commuting covariant constant
spinors $\lambda_\alpha (x^i)$ and $\bar\lambda_{\dot\lambda} (x^i)$ are
fixed on the configuration space, and an the other hand, time-like
depending Grassmann variables are not spinors. Then the time-like
components of the Rarita-Schwinger field may be written as

\be
\psi_0^\alpha (x^i, t)=\lambda^\alpha (x^i) \psi (t) .
\en

The spatial components of the Rarita-Schwinger field have the following
representation corresponding to the direct product time-subspace on the
3-space of the fixed spatial configuration (in our case it is a plane or
a three sphere). Explicitly, we get 

\be
\psi^\alpha_m (x^i, t)= e^{(\mu )}_m \sigma^{\alpha\dot\beta}_{(\mu )}
\bar\lambda_{\dot\beta} (x^i) \bar\lambda (t) ,
\en

\noi where $e^{(\mu )}_m (x^i,t)$ are the tetrads for the FRW metric. Those
representations are solutions of the supergravity equations.

\bi

We have the classical canonical Hamiltonian

\be 
H_{can} = N H + \frac{1}{2} \bar\psi S - \frac{1}{2} \psi \bar S
+ \frac {1}{2} V {\cal F} , 
\en

\noi where $H$ is the Hamiltonian of the system, $S$ and $\bar S$ are
supercharges and ${\cal F}$ is the ${\cal U}(1)$ rotation generator. The
form of the canonical Hamiltonian (27) explains the fact, that $N, \psi ,
\bar\psi$ and $V$ are Lagrange multipliers which enforce only the
first-class constraints, $H = 0, S = 0, \bar S = 0$ and ${\cal F} = 0,$
which express the invariance of the conformal $n = 2$ supersymmetric
transformations. As usual with the Grassmann variables we have the
second-class constraints, which can be eliminated by the Dirac procedure.
In the usual canonical quantization the even canonical variables change by
operators

\be
R \to R,\,\,\, \pi_R = i\frac{\partial}{\partial R}\quad ;\quad Z^A \to Z^A
,\quad \pi_A = i\frac{\partial}{\partial Z^A}
\en  

\noi and the odd variables $\lambda , \bar\lambda , \chi^A , \bar\chi^A ,
\phi^A$ and $\bar\phi^A$ after quantization become anticonmutators.

\bi

We can write  $\lambda , \bar\lambda , \chi_A,  \bar\chi_A , \phi_A$ and
$\bar\phi_A$ in the form of the direct product $1+2 n , 2\times 2$ 
matrices.
We then obtain a matrix realization for the case of  $n$ complex mater
supermultiplets

\[
\lambda = \sqrt{\frac{3}{2}} \sigma_1^{(-)} \otimes 1_2 \otimes\ldots
\otimes 1_{2n+1} , 
\]
\be
\lambda^+ = \sqrt{\frac{3}{2}} \sigma_1^{(+)}
\otimes 1_2\otimes\ldots\otimes 1_{2n+1},  
\en
\[
\phi^A = \kappa \sigma^{(3)}_1 \otimes..\otimes \sigma^{(3)}_{2A-1}
\otimes\sigma_{2A}^{(-)} \otimes 1_{2A+1}\otimes.. 
1_{2n+1} ,
\]
\[
\bar\phi_A = \kappa \sigma_1^{(3)} \otimes..\otimes
 \sigma^{(3)}_{2A-1} \sigma \otimes \sigma^{(+)}_{2A}\otimes
1_{2A+1}..\otimes 1_{2n+1} ,
\]
\[
\chi^A =  \kappa\sigma^{(3)}_1 \otimes..\otimes \sigma^{(3)}_{2A}
\otimes \sigma^{(-)}_{(2A+1)} \otimes 1_{2A+1}..\otimes 
1_{2n+1} , 
\]
\[
\bar\chi^A = \kappa \sigma_1^{(3)} \otimes..\otimes\sigma^{(3)}_{2A}
\otimes \sigma^{(+)}_{2A+1} \otimes 1_{2A+2}
..\otimes 1_{2n+1},  
\]

\noi where the down index in the direct product at the matrix shows the
place of the matrix $(A = 1,2, \ldots ,n), \sigma^\pm =
\frac{\sigma^{1}\pm i\sigma^2}{2}$ with $\sigma^1 , \sigma^2 ,$ and
$\sigma^3$ being the Pauli Matrices. 

In the matrix realization the operators $\lambda , \chi^A$ and
$\phi^A$ on the wave function $\psi = \psi (R,Z^A, \bar Z^A, \bar\lambda ,
\bar\chi , \bar\phi)$ are $2^{2n+1}$ component columns
$\psi_i (R,Z^A, \bar Z^A) ,$
$ (i = 1, \ldots , 2^{2n+1}).$
In the quantum theory the first class constraints associated with the
invariance of action (18) become conditions on the wave functions 
$\psi$. Therefore any physically allowed states must obey the quantum
constraints

\be
H \psi = 0, S\psi = 0, \bar S\psi = 0, {\cal F} \psi  = 0 , 
\en

Where the first equation in (30) is the so-called Wheeler De Witt equation
for minisuperspace models.

\bi

To obtain the quantum expression for the Hamiltonian H and for the
supercharges ${\cal S}$ and $S^+$ we must solve the operator ordering
ambiguity. Such ambiguities always arise when, as in our case, the
operator expression contains the product of non-commuting operators $R,
\pi_R, Z^A$ and $\pi_A$. Then we must integrate with measure
$R^{\frac{1}{2}} (\mbox{det} G_{\bar A})^{1/2} d R d^A_z d^A_{\bar Z}$ in
the inner product of two states. In this measure the momenta $\pi_R
=i\frac{\partial}{\partial R}$ is non-Hermitian with $\pi^+_R = R^{-1/2}
\pi_R R^{\frac{1}{2}}$; however, the combination $(R^{-1/2} \pi_R)^+ =
\pi^+_R R^{-1/2} = R^{-1/2} \pi_R$ is Hermitian. The canonical momenta
$\pi^+_A$, Hermitian-conjugate to $\pi_A =i\frac{\partial}{\partial Z^A}$,
have the form $(\pi_A)^+ = g^{-1/2}(\bar\pi_{\bar A}) g^{\frac{1}{2}}$,
where $g=\mbox{det} \, G_{A\bar B}$. 

\bi

The quantum generators $H,S,\bar S$ and ${\cal F}$ form a closed 
superalgebra of the supersymmetric quantum mechanics

\[
\{ S,\bar S \} = 2H, \quad [S,H]=[\bar S, H]= 0 , 
\quad S^2 =\bar S^2 = 0 
\]
\be
\left[ {\cal F}, S \right] = -S, \quad [ {\cal F},\bar S ] =\bar S, 
\quad [ {\cal F}, H]= 0 . 
\en  

As we can see from Hamiltonian, the energy of the scale factor is negative.
This is reflected in the fact that the anticommutator value 
$\{\lambda ,\bar\lambda \} =-3/2$ of superpartners $\lambda$ and 
$\bar\lambda$ of the scale factor is negative, unlike anticommutation 
relations for $\chi^A$, $\bar\chi_B$ and $\phi^A , \bar\phi_B$, which are
positive. Anticommutation relations may be satisfied under the
conditions.
\be
\bar\lambda =- \lambda^+, (\chi^A)^+ = \bar\chi_A, (\phi_A)^+ =\bar\phi_A ,
\en

where $\{\lambda , \lambda^+\} =\frac{3}{2}$. Then the equation may be 
written in the form

\be
\bar\lambda = \xi^{-1} \lambda^+ \xi , \bar\chi_A = \xi^{-1} (\chi^A)^+
\xi , \bar\phi_A =\xi^{-1} (\phi^A)^+ \xi .
\en

In order to have consistency with expressions (32) and (33) it is necessary
that the operator $\xi$ possess the following properties $(\xi^+ = \xi )$:

\[
\lambda^+\xi =-\xi\lambda^+ , (\chi^A)^+\xi=\xi (\chi^A)^+, 
\]
\be
(\phi^A)^+ \xi
=\xi (\phi^A)^+ .
\en

The operator $\bar\lambda , \bar\chi_A$ and $\bar\phi_A$ will be conjugate
to operators $\lambda ,\chi^A$ and $\phi^A$ under inner product of two
states $\psi_1$ and $\psi_2$

\be
< \psi_1,\psi_2>_q =\int \psi^*_1 |\xi | \psi_2 R^{1/2} g^{1/2} dR 
d^n \bar z d^n \bar z , 
\en

\noi which in general is non-positive. In the matrix realization the 
operator $\xi$ has the form 
\be
\xi = \sigma^{(3)}_1 \otimes 1_2 \otimes \ldots \otimes 
1_{2n+1}
\en

\bi

So, for the superchange operator $S$ we can construct conjugation (33)
under the operator $\bar S$ with the help of the following equation
\be
\bar S = \xi^{-1} S^+ \xi .
\en

We can see that the anticommutators of supercharge $S$ and their conjugate
$\bar S$ under our conjugate operation has the form 
\be
\{ \overline{S, \bar S} \} = \xi^{-1} \{ S,\bar S^+\} \xi =
\{ S,\bar S \}
\en 

\noi and it is self-conjugate operator.

\bi

As a consequence of algebra (31) we obtain that the Hamiltonian $H$ is a
self-conjugate operator $\bar H = \xi^{-1} H^+ \xi =H$ and its value is
real.

\bi

Note that the superalgebra (31) does not define positive-definite Hamiltonian
in a full agreement with the circunstance, that the potential 
$V_{eff} (z,\bar z)$ of scalar fields (20,21) is not positive 
semi-definite in contrast with the standard supersymmetric quantum 
mechanics.\\
In this case the normalizable solution to the quantum constraints

\be
S\psi = 0 ,  \quad \bar S \psi = 0
\en

\noi is the wavefunction in the supersymmetry breaking state with 
zero energy.\\
With the conformal algebra given by (31) we need to solve only these
two quantum constraints in order to search our solutions. Using the
matrix representation (29) to solve (39) one $\psi_{2^{2n+1}}$ component
$\psi$ can have the right behaviour when $R\to \infty$, we have
a normalizable solution.

\be
\psi_{(R, z,\bar z)} = CR^{3/4} e^{-\frac{1}{2} (1-T_3) 
(2 \frac{e^{G/2}}{\kappa^3}R^3 -3\sqrt{k} \frac{R^2}{\kappa^2})} 
\psi_0
\en
 
\noi where $T_3 =\frac{1}{2} (\sigma^{(3)}_1 + \sigma^{(3)}_{1+2n}),
\quad \psi_0 = \left( 
\begin{array}{c}
1 \\ 
\vdots \\ 
1 
\end{array} \right)$ and 
\[
\sigma^{(3)}_1 = \sigma^3_1 \otimes 1_{(2)} \otimes
\ldots \otimes 1_{1+2n}, 
\]
\be
\sigma^{(3)}_{1+2n} = 1_1 \otimes \ldots \otimes 1_{2n} \otimes
\sigma^3_{1+2n} 
\en

\bi

In the case of a minimum the potential $V_{eff} (z_0, \bar z_0)=0$ and
$k=0$, then using (29) we get 

\be
\psi_{2^{1+2n}}  (R) = \tilde C_0 R^{3/4} e^{-2m_{3/2} M^2_{pl} R^3} 
\en

\noi where we have thus

\be
1 = \tilde C_0 \int^\infty_0 R^{3/2} e^{-4m_{3/2} M^2 pl R^3}
R^{1/2} dR ,
\en

\noi the normalization constant has the following value $\tilde C_0 =
(12m_{3/2} M^2_{pl})^{\frac{1}{2}}$.
\be
\rho(R)dR \equiv |\psi_{2^{2l+1}}|^2R^{1/2}dR,
\en
which give us the probability to find the Universe with scale factor 
between $R$ and $R + dR$, as usual in quantum mechanics. Then, the 
probability (also called distribution function) of having a Universe with 
scale factor $R$ is 
\begin{eqnarray}
P(R) &=& \int^R_0|\psi_{2^{2l+1}}|^2R^{1/2}dR,\nonumber\\
&=& 1 - e^{-4m_{3/2}M_{pl}^2R^3}.
\end{eqnarray}

\section{Conclusions}
The specific quantum supersymmetric mechanics corresponding to quantum
level in our models defines the structure which permits the fundamental
states invariant under the $n=2$ local conformal supersymmetry in $N=1$
supergravity interacting with a set of matter fields \cite{8}. In our
case the constraints and the wave function of the universe permit the
existence of non-trivial solutions.
\bi

\Acknow{This work was partially suported by CONACyT, grant No. 28454E.}


\newpage

\small
\begin{thebibliography} {99}
\bibitem{1} B.S. De Witt, {\it Phys. Rev.\/} 160, 1143 (1967); J.A. Wheeler,
             `` Relativity Groups and Topology'', eds. C. De Witt and 
             B. De Witt, Gordon and Breach, 1969; M.P. Ryan, 
              ``Hamiltonian Cosmology'', Springer-Verlag, 1971.
\bibitem{2} A. Mac\ii as, O. Obreg\'on and M.P. Ryan, Jr., {\it Class 
             Quantum Grav.} {\bf 4}, 1477 (1987); P.D. D'Eath and D.I. 
             Hughes, {\it Phys. Lett}. {\bf B214}, 498 (1988).
\bibitem{3} R. Graham, {\it Phy. Rev. Lett}. {\bf 67}, 1381 (1991).
\bibitem{4} P.D. D'Eath, ``Supersymmetric Quantum Cosmology'', Cambridge
              Univ. Press, 1996. 
\bibitem{5} P.V. Moniz, {\it Int. J. Mod. Phys.} {\bf A11},4321 (1996).  
\bibitem{6} O. Obreg\'on, J.J. Rosales and V.I. Tkach, {\it Phys. Rev.}
           {\bf D53}, R1750 (1996); V.I. Tkach, J.J. Rosales and O.
           Obreg\'on, {\it Class. Quantum Grav.} 13, 2349 (1996).
\bibitem{7} V.I. Tkach, O. Obreg\'on and J.J. Rosales, {\it Class. Quantum 
           Grav.} {\bf 14}, 339 (1997); V.I. Tkach, J.J. Rosales and 
           J. Mart\ii nez, {\it Class. Quantum Grav.}, {\bf 15}, 3755 
           (1998); V.I. Tkach,
           J.J. Rosales and J. Socorro, {\it Class. Quantum Grav.} {\bf 16}
           797 (1999); O. Obre\'on, J.J. Rosales, J. Socorro and V.I. Tkach,
           {\it Class. Quantum Grav.}{\bf 16},2861 (1999).
\bibitem{8} E. Cremmer, B. Julia, J. Scherk, S. Ferrara, L. Girardello 
            and P. van Niewenhuizen, {\it Nucl. Phys.} {\bf B147}, 105 (1979);
            V.S. Kaplunovsky and J. Louis, {\it Phys. Lett.} B306. 269 
            (1993); P. Nath, R. Arnowitt and C. Chamseddine, ``Applied
            $N=1$ Supergravity'', World Scientific, 1984.
\bibitem{9} H.B. Hartle and S.W. Hawking, {\it Phys. Rev.} {\bf D28}, 
               2960 (1983); A.D. Linde, {\it Sov. Phys. JETP} {\bf 60},
               211 (1984).
\bibitem{10} F. Cooper, A. Khare and U. Sukhatme, {\it Phys. Rep.} 
              {\bf 251}, 267 (1995); C.V. Suluman, {\it J. Phys.} {\bf A18},
              2917 (1985); P. Salomonson and J. W. Van Holten, {\it Nucl.
              Phys.} {\bf B196} 509 (1982).
\bibitem{11} E. Witten, {\it Nucl. Phys. B} 186, 412 (1981); C. 
               Randjbar-Doemi, A. Salam and J. Strathdee, {\it Nucl. Phys. B}
               214, 419 (1983). D.P. Sorokin, V.I. Tkach and D.V. Volkov, 
               {\it Phys. Lett. B} 161, 301 (1985).

\end{thebibliography}
\end{document}